\documentclass[amstex,12pt]{article}
\usepackage{amssymb}
\newtheorem{theorem}{Theorem}
\usepackage{amsmath}
\usepackage{epsfig}

\begin{document}

\title{Poisson-Vlasov in a strong magnetic field: A stochastic solution approach}
\author{R. Vilela Mendes\thanks{%
IPFN - EURATOM/IST Association, Instituto Superior T\'{e}cnico, Av. Rovisco
Pais 1, 1049-001 Lisboa, Portugal} \thanks{%
CMAF, Complexo Interdisciplinar, Universidade de Lisboa, Av. Gama Pinto, 2 -
1649-003 Lisboa (Portugal), e-mail: vilela@cii.fc.ul.pt,
http://label2.ist.utl.pt/vilela/}}
\date{}
\maketitle

\begin{abstract}
Stochastic solutions are obtained for the Maxwell-Vlasov equation in the
approximation where magnetic field fluctuations are neglected and the
electrostatic potential is used to compute the electric field. This is a
reasonable approximation for plasmas in a strong external magnetic field.
Both Fourier and configuration space solutions are constructed.


\end{abstract}

\section{Introduction. The notion of stochastic solution}

The solutions of linear elliptic and parabolic equations, with Cauchy or
Dirichlet boundary conditions, have a probabilistic interpretation. These
are classical results which may be traced back to the work of Courant,
Friedrichs and Lewy \cite{Courant} in the 1920's and became a standard tool
in potential theory\cite{Getoor} \cite{Bass1} \cite{Bass2}. A simple example
is provided by the heat equation

\begin{equation}
\partial _{t}u(t,x)=\frac{1}{2}\frac{\partial ^{2}}{\partial x^{2}}%
u(t,x)\qquad \mathnormal{with}\qquad u(0,x)=f(x)  \label{1.1}
\end{equation}
with solution written either as 
\begin{equation}
u\left( t,x\right) =\frac{1}{\sqrt{2\pi }}\int \frac{1}{\sqrt{t}}\exp \left(
-\frac{\left( x-y\right) ^{2}}{2t}\right) f\left( y\right) dy  \label{1.2}
\end{equation}
or as 
\begin{equation}
u(t,x)=\mathbb{E}_{x}f(X_{t})  \label{1.3}
\end{equation}
$\mathbb{E}_{x}$ denoting the expectation value, starting from $x$, of the
process 
\[
dX_{t}=dW_{t} 
\]
$W_{t}$ being the Wiener process.

Eq.(\ref{1.1}) is a \textit{specification} of a problem whereas (\ref{1.2})
and (\ref{1.3}) are \textit{solutions} in the sense that they both provide
algorithmic means to construct a function satisfying the specification. An
important condition for (\ref{1.2}) and (\ref{1.3}) to be considered as
solutions is the fact that the algorithms are independent of the particular
solution, in the first case an integration procedure and in the second the
simulation of a solution-independent process. In both cases the algorithm is
the same for all initial conditions. This should be contrasted with
stochastic processes constructed from particular solutions, as has been done
for example for the Boltzman equation\cite{Graham}.

In contrast with the linear problems, explicit solutions for nonlinear
partial differential equations, in terms of elementary functions or
integrals, are only known in very particular cases. Hence it is in the field
of nonlinear partial differential equations that the stochastic method might
be most useful. Whenever a solution-independent stochastic process is found
that, for arbitrary initial conditions, generates the solution in the sense
of Eq.(\ref{1.3}), an exact stochastic solution is obtained. In this way the
set of equations for which exact solutions are known might be considerably
extended.

The exit measures provided by diffusion plus branching processes\cite
{McKean3} \cite{Dynkin1} \cite{Dynkin4} \cite{Dynkin3} \cite{Dynkin2} \cite
{Dynkin5} as well as the stochastic representations recently constructed for
the Navier-Stokes\cite{Jan} \cite{Waymire} \cite{Waymire2} \cite{Bhatta1} 
\cite{Ossiander} \cite{Orum}, the Poisson-Vlasov\cite{Vilela1} \cite{Vilela2}%
, the Euler\cite{Vilela3} and a nonlinear fractional differential equation%
\cite{Cipriano} define initial condition-independent processes for which the
mean values of some functionals are solutions to these equations. Therefore,
they are exact \textbf{stochastic solutions}.

Typically, in the stochastic solutions, one deals with a process that starts
from the point where the solution is to be found, the solution being then
obtained from a functional computed along the whole sample path or until the
process hits the boundary. In addition to providing new exact results, the
stochastic solutions are also a promising tool for numerical implementation.
Here the relevant question is to know when a stochastic algorithm is
competitive with the existing deterministic algorithms. Although there is no
general answer to this question, there are a few considerations that suggest
where and when stochastic algorithms might be useful, namely:

(i) Deterministic algorithms grow exponentially with the dimension $d$ of
the space, roughly $N^{d}$ ($\frac{L}{N}$ being the linear size of the
grid). This implies that to have reasonable computing times, the number of
grid points may not be sufficient to obtain a good local resolution for the
solution. In contrast a stochastic simulation only grows with the dimension
of the process, typically of order $d$.

(ii) In general, deterministic algorithms aim at obtaining the behavior of
the solution in the whole domain. That means that, even if an efficient
deterministic algorithm exists for the problem, a stochastic algorithm might
still be competitive if only localized values of the solution are desired.
This comes from the very nature of the stochastic representation processes
that always starts from a definite point in the domain. According to what is
desired, configuration or Fourier space representations should be used. For
example by studying only a few high Fourier modes one may obtain information
on the small scale fluctuations that only a very fine grid would provide in
a deterministic algorithm.

(iii) Each time a sample path of the process is implemented, it is
independent from any other sample paths that are used to obtain the
expectation value. Likewise, paths starting from different points are
independent from each other. Therefore the stochastic algorithms are a
natural choice for parallel and distributed implementation. Provided some
differentiability conditions are satisfied, the process also handles equally
well simple or complex boundary conditions.

(iv) Stochastic algorithms may also be used for domain decomposition
purposes \cite{Acebron1} \cite{Acebron2} \cite{Talay}. One may, for example,
decompose the space in subdomains and then use in each one a deterministic
algorithm with Dirichlet boundary conditions, the values on the boundaries
being determined by a stochastic algorithm.

Stochastic solutions also provide an intuitive characterization of the
physical phenomena, relating nonlinear interactions to cascading processes.
By the study of exit times from a domain they also provide access to
quantities that cannot be obtained by perturbative methods\cite{Eleuterio} 
\cite{VilelaZeit}.

One way to construct stochastic solutions is based on a probabilistic
interpretation of the Picard series. First, the differential equation is
written as an integral equation. Then there are two possibilities.

In the first case the series is rearranged in a such a way that the
coefficients of the successive terms in the Picard iteration, including the
initial condition term, obey a normalization condition. The stochastic
solution is then equivalent to importance sampling of the normalized Picard
series. In this case the stochastic process, that constructs the solution,
is a branching process, the branching being controlled by the nonlinear part
of the equation.

The second possibility occurs when the initial condition term is not
multiplied by a probability factor. Then, the integral of the integral
equation may still be given a probabilistic interpretation but the process
that is used for the construction of the solution is more general
tree-indexed stochastic process.

In this paper, pursuing the work on kinetic equations initiated in \cite
{Vilela1} and \cite{Vilela2}, solutions are obtained for the Maxwell-Vlasov
equation in the approximation where magnetic field fluctuations are
neglected and the electrostatic potential is used to compute the electric
field. This is a reasonable approximation for plasmas in a strong external
magnetic field. Both Fourier and configuration space solutions are
constructed.

In Sect.2.1 one discusses the formulation of the full Maxwell-Vlasov system
as an integral equation for the charge densities. In Sects. 2.1 to 2.4 the
solutions to the Fourier-transformed equation are obtained both for a static
uniform and a slowly varying magnetic field. The stochastic processes
associated to the construction of these solutions are branching processes
with the waiting time controlled by the velocity Fourier component and the
densities (anti-) evolved by the linear part of the equation.

In Sect.3 one deals with the configuration space equation and in this case
the most natural stochastic formulation involves a general tree-indexed
stochastic process.

\section{The Poisson-Vlasov equation in a magnetic field}

\subsection{The Maxwell-Vlasov system}

Consider a two-species Maxwell-Vlasov system in 3+1 space-time dimensions 
\begin{equation}
\frac{\partial f_{i}}{\partial t}+\stackrel{\rightarrow }{v}\cdot \nabla
_{x}f_{i}+\frac{e_{i}}{m_{i}}\left( \stackrel{\rightarrow }{E}+\frac{%
\stackrel{\rightarrow }{v}}{c}\times \stackrel{\rightarrow }{B}\right) \cdot
\nabla _{v}f_{i}=0  \label{4.1}
\end{equation}
$\left( i=1,2\right) $, with 
\begin{eqnarray}
\frac{\partial }{\partial t}\stackrel{\rightarrow }{E}-c\nabla _{x}\times 
\stackrel{\rightarrow }{B} &=&-4\pi \sum_{i}e_{i}\int \stackrel{\rightarrow 
}{v}f_{i}d^{3}v  \nonumber \\
\frac{\partial }{\partial t}\stackrel{\rightarrow }{B}+c\nabla _{x}\times 
\stackrel{\rightarrow }{E} &=&0  \nonumber \\
\nabla _{x}\cdot \stackrel{\rightarrow }{E} &=&4\pi \sum_{i}e_{i}\int
f_{i}d^{3}v  \nonumber \\
\nabla _{x}\cdot \stackrel{\rightarrow }{B} &=&0  \label{4.2}
\end{eqnarray}
To study the Vlasov-Maxwell system as a nonlinear equation for the $%
f_{i}\left( t,x,v\right) $ densities one has to obtain explicit expressions
for the electromagnetic fields in terms of the charge densities. Define
scalar and vector potentials 
\begin{eqnarray}
\stackrel{\rightarrow }{E} &=&-\nabla \phi -\frac{1}{c}\frac{\partial 
\stackrel{\rightarrow }{A}}{\partial t}  \nonumber \\
\stackrel{\rightarrow }{B} &=&\nabla \times \stackrel{\rightarrow }{A}
\label{4.3}
\end{eqnarray}
which, in the Lorentz gauge ($\nabla \cdot \stackrel{\rightarrow }{A}+\frac{1%
}{c}\frac{\partial \phi }{\partial t}=0$), obey the equations 
\begin{eqnarray}
\Delta \phi -\frac{1}{c^{2}}\frac{\partial ^{2}\phi }{\partial t^{2}}
&=&-4\pi \sum_{i}e_{i}\int f_{i}d^{3}v  \nonumber \\
\Delta \stackrel{\rightarrow }{A}-\frac{1}{c^{2}}\frac{\partial ^{2}%
\stackrel{\rightarrow }{A}}{\partial t^{2}} &=&-\frac{4\pi }{c}%
\sum_{i}e_{i}\int \stackrel{\rightarrow }{v}f_{i}d^{3}v  \label{4.4}
\end{eqnarray}
Using the (retarded) Green's function for the wave equation, 
\begin{equation}
G\left( \stackrel{\rightarrow }{x}t,\stackrel{\rightarrow }{x}^{\prime
}t^{^{\prime }}\right) =\frac{1}{\left| \stackrel{\rightarrow }{x}-\stackrel{%
\rightarrow }{x}^{\prime }\right| }\delta \left( t^{\prime }+\frac{\left| 
\stackrel{\rightarrow }{x}-\stackrel{\rightarrow }{x}^{\prime }\right| }{c}%
-t\right)  \label{4.5}
\end{equation}
one obtains 
\begin{eqnarray}
\phi \left( \stackrel{\rightarrow }{x},t\right) &=&\int d^{3}x^{\prime }%
\frac{1}{\left| \stackrel{\rightarrow }{x}-\stackrel{\rightarrow }{x}%
^{\prime }\right| }\sum_{i}e_{i}\int f_{i}\left( \stackrel{\rightarrow }{x}%
^{\prime },\stackrel{\rightarrow }{v},t-\frac{\left| \stackrel{\rightarrow }{%
x}-\stackrel{\rightarrow }{x}^{\prime }\right| }{c}\right) d^{3}v  \nonumber
\\
\stackrel{\rightarrow }{A}\left( \stackrel{\rightarrow }{x},t\right) &=&\int
d^{3}x^{\prime }\frac{1}{\left| \stackrel{\rightarrow }{x}-\stackrel{%
\rightarrow }{x}^{\prime }\right| }\sum_{i}\frac{e_{i}}{c}\int \stackrel{%
\rightarrow }{v}f_{i}\left( \stackrel{\rightarrow }{x}^{\prime },\stackrel{%
\rightarrow }{v},t-\frac{\left| \stackrel{\rightarrow }{x}-\stackrel{%
\rightarrow }{x}^{\prime }\right| }{c}\right) d^{3}v  \label{4.6}
\end{eqnarray}
In writing (\ref{4.6}) as a source term solution of (\ref{4.4}), one has
assumed that the initial conditions for the equations (\ref{4.4}) vanish
together with their time derivatives or, alternatively, that the initial
time is in the remote past so that there are no more contributions from the
initial conditions. For a more general solution which should be used for
transitory phenomena or plasma probing by short localized pulses see \cite
{Initial}.

Use of (\ref{4.3}) yields 
\begin{equation}
\begin{array}{lll}
\stackrel{\rightarrow }{E} & = & \int d^{3}x^{^{\prime }}\sum_{i}e_{i}\frac{1%
}{\left| x-x^{^{\prime }}\right| }\int d^{3}v\left\{ \frac{\stackrel{%
\longrightarrow }{x-x^{^{\prime }}}}{\left| x-x^{^{\prime }}\right| ^{2}}%
+\left( \frac{\stackrel{\longrightarrow }{x-x^{^{\prime }}}}{c\left| 
\stackrel{\rightarrow }{x}-\stackrel{\rightarrow }{x}^{\prime }\right| }-%
\frac{\stackrel{\longrightarrow }{v}}{c^{2}}\right) \partial _{t}\right\}
f\left( x^{^{\prime }},v,t-\frac{\left| x-x^{^{\prime }}\right| }{c}\right)
\\ 
\stackrel{\rightarrow }{B} & = & \int d^{3}x^{^{\prime }}\sum_{i}\frac{e_{i}%
}{c}\frac{1}{\left| x-x^{^{\prime }}\right| }\int d^{3}vv\times \left\{ 
\frac{\stackrel{\longrightarrow }{x-x^{^{\prime }}}}{\left| x-x^{^{\prime
}}\right| ^{2}}+\frac{\stackrel{\longrightarrow }{x-x^{^{\prime }}}}{c\left| 
\stackrel{\rightarrow }{x}-\stackrel{\rightarrow }{x}^{\prime }\right| }%
\partial _{t}\right\} f\left( x^{^{\prime }},v,t-\frac{\left| x-x^{^{\prime
}}\right| }{c}\right)
\end{array}
\label{4.7}
\end{equation}
Replacing now Eq.(\ref{4.7}) in (\ref{4.1}) one sees that, as a function of
the densities $f\left( x,v,t\right) $, Maxwell-Vlasov is a nonlinear and
nonlocal (in time) differential equation. Its full stochastic solution
treatment will be dealt with elsewhere. Here one deals with approximations
of practical interest for fusion plasmas. First notice that for
non-relativistic plasmas the last terms in (\ref{4.7}) are small and in the
quasi-static approximation one has 
\begin{equation}
\begin{array}{lll}
\stackrel{\rightarrow }{E} & = & \int d^{3}x^{^{\prime }}\sum_{i}e_{i}\frac{%
\stackrel{\longrightarrow }{x-x^{^{\prime }}}}{\left| x-x^{^{\prime
}}\right| ^{3}}\int d^{3}vf\left( x^{^{\prime }},v,t\right) \\ 
\stackrel{\rightarrow }{B} & = & \int d^{3}x^{^{\prime }}\sum_{i}\frac{e_{i}%
}{c}\frac{1}{\left| x-x^{^{\prime }}\right| ^{3}}\int d^{3}v\left( \stackrel{%
\longrightarrow }{v}\times \left( \stackrel{\longrightarrow }{x-x^{^{\prime
}}}\right) \right) f\left( x^{^{\prime }},v,t\right)
\end{array}
\label{4.8}
\end{equation}
Furthermore, for microturbulence studies in fusion plasmas in strong
magnetic fields, a reasonable approximation neglects magnetic field
fluctuations and uses the electrostatic potential of the charges to compute
the electric field. This is what will be called Poisson-Vlasov in a static
(external) magnetic field.

\subsection{Poisson-Vlasov in a static magnetic field}

In this approximation the equation is 
\begin{eqnarray}
0 &=&\frac{\partial f_{i}}{\partial t}+\left( \stackrel{\rightarrow }{v}%
\cdot \nabla _{x}+\frac{e_{i}}{m_{i}}\frac{\stackrel{\rightarrow }{v}}{c}%
\times \stackrel{\rightarrow }{B}\left( x\right) \cdot \nabla _{v}\right)
f_{i}  \nonumber \\
&&+\frac{e_{i}}{m_{i}}\int d^{3}x^{^{\prime }}\sum_{j}e_{j}\int
d^{3}uf_{j}\left( x^{^{\prime }},u,t\right) \frac{\stackrel{\longrightarrow 
}{x-x^{^{\prime }}}}{\left| x-x^{^{\prime }}\right| ^{3}}\cdot \nabla
_{v}f_{i}\left( x,v,t\right)  \label{3.1}
\end{eqnarray}
or, in the Fourier transformed version 
\begin{equation}
F\left( \xi ,t\right) =\frac{1}{\left( 2\pi \right) ^{3}}\int d^{6}\eta
f\left( \eta ,t\right) e^{i\xi \cdot \eta }  \label{3.3}
\end{equation}
with $\eta =\left( \stackrel{\rightarrow }{x},\stackrel{\rightarrow }{v}%
\right) $ and $\xi =\left( \stackrel{\rightarrow }{\xi _{1}},\stackrel{%
\rightarrow }{\xi _{2}}\right) \circeq \left( \xi _{1},\xi _{2}\right) $, 
\begin{equation}
\begin{array}{ll}
\frac{\partial F_{i}\left( \xi ,t\right) }{\partial t} & =\left( \stackrel{%
\rightarrow }{\xi _{1}}\cdot \nabla _{\xi _{2}}+\frac{e_{i}}{cm_{i}}\nabla
_{\xi _{2}}\times \stackrel{\rightarrow }{B}\left( -i\nabla _{\xi
_{1}}\right) \cdot \stackrel{\rightarrow }{\xi _{2}}\right) F_{i}\left( \xi
,t\right) \\ 
& -\frac{4\pi e_{i}}{m_{i}}\int d^{3}\xi _{1}^{^{\prime }}F_{i}\left( \xi
_{1}-\xi _{1}^{^{\prime }},\xi _{2},t\right) \frac{\stackrel{\rightarrow }{%
\xi _{2}}\cdot \stackrel{\rightarrow }{\xi _{1}^{^{\prime }}}}{\left| \xi
_{1}^{^{\prime }}\right| ^{2}}\sum_{j}e_{j}F_{j}\left( \xi _{1}^{^{\prime
}},0,t\right)
\end{array}
\label{3.4}
\end{equation}

The aim is to obtain stochastic solutions both for the equation in
configuration space and for its Fourier-transformed version. Because of the
localized nature of the stochastic solutions, as discussed in the
introduction, both types of solutions are useful for the applications. If,
in a plasma confinement device, one is interested in the behavior of the
solution at a particular point (for example at a point either in the core or
in the scrape-off layer) then it is the solution in configuration space that
is useful. If however one is interested in the nature of the turbulent
fluctuations it is probably the study of high Fourier modes in the
Fourier-transformed equation that will be most useful.

Eqs.(\ref{3.1}) and (\ref{3.4}) have linear and nonlinear parts. The linear
evolutions are, respectively 
\begin{equation}
\begin{array}{lll}
f_{i}^{(0)}\left( x,v,t\right) & = & e^{-tQ_{\eta }}f_{i}^{(0)}\left(
x,v,0\right) \\ 
F_{i}^{(0)}\left( \xi _{1},\xi _{2},t\right) & = & e^{-tQ_{\xi
}}F_{i}^{(0)}\left( \xi _{1},\xi _{2},0\right)
\end{array}
\label{3.4a}
\end{equation}
the operators $Q_{\eta }$ and $Q_{\xi }$ being 
\begin{equation}
\begin{array}{lll}
Q_{\eta } & = & \stackrel{\rightarrow }{v}\cdot \nabla _{x}+\frac{e_{i}}{%
cm_{i}}\stackrel{\rightarrow }{v}\times \stackrel{\rightarrow }{B}\left(
x\right) \cdot \nabla _{v} \\ 
Q_{\xi } & = & -\stackrel{\rightarrow }{\xi _{1}}\cdot \nabla _{\xi _{2}}-%
\frac{e_{i}}{cm_{i}}\nabla _{\xi _{2}}\times \stackrel{\rightarrow }{B}%
\left( -i\nabla _{\xi _{1}}\right) \cdot \stackrel{\rightarrow }{\xi _{2}}
\end{array}
\label{3.4b}
\end{equation}
From (\ref{3.4a}) and (\ref{3.4b}) it follows that the linear evolution of
the function arguments $x$, $v$, $\xi _{1}$ and $\xi _{2}$ is ruled by the
following equations 
\begin{equation}
\begin{array}{lll}
\frac{d}{dt}x\left( t\right) & = & -v\left( t\right) \\ 
\frac{d}{dt}v\left( t\right) & = & -\frac{e_{i}}{cm_{i}}\left( v\left(
t\right) \times B\left( x\left( t\right) \right) \right)
\end{array}
\label{3.4d}
\end{equation}
\begin{equation}
\begin{array}{lll}
\frac{d}{dt}\xi _{1}\left( t\right) & = & -\frac{e_{i}}{cm_{i}}\left( \nabla
_{\xi _{2}}\left( t\right) \times i\nabla B\left( -i\nabla _{\xi _{1}}\left(
t\right) \right) \cdot \xi _{2}\left( t\right) \right) \\ 
\frac{d}{dt}\xi _{2}\left( t\right) & = & \xi _{1}\left( t\right) +\frac{%
e_{i}}{cm_{i}}B\left( -i\nabla _{\xi _{1}}\left( t\right) \right) \times \xi
_{2}\left( t\right)
\end{array}
\label{3.4e}
\end{equation}
and 
\begin{equation}
\begin{array}{lll}
\frac{d}{dt}\nabla _{\xi _{1}}\left( t\right) & = & -\nabla _{\xi
_{2}}\left( t\right) \\ 
\frac{d}{dt}\nabla _{\xi _{2}}\left( t\right) & = & -\frac{e_{i}}{cm_{i}}%
\left( \nabla _{\xi _{2}}\left( t\right) \times B\left( -i\nabla _{\xi
_{1}}\left( t\right) \right) \right)
\end{array}
\label{3.4f}
\end{equation}

One sees from (\ref{3.4d}) that the linear evolution of the densities $%
f\left( x,v,t\right) $ in configuration space acts only on the arguments of
the function. However, from (\ref{3.4e}) and (\ref{3.4f}) one also sees
that, if the magnetic field $B$ is not constant in space, the linear
evolution of the Fourier transformed densities $F_{i}^{(0)}\left( \xi
_{1},\xi _{2},t\right) $ is more complex, involving derivatives of the
Fourier density. For the stochastic solutions, one associates a process to
each function, therefore it is not convenient to use the full linear part of
the evolution operator in the Fourier-transformed equation. Such problem
does not exist if the static magnetic field is also uniform in space. Here
one starts by studying this case, which is then extended to the case of a
slowly varying magnetic field.

\subsection{Fourier-transformed Poisson-Vlasov in a static uniform magnetic
field}

Consider a uniform magnetic field $\stackrel{\rightarrow }{B}=\stackrel{%
\rightarrow }{B}_{0}=B_{0}\stackrel{\wedge }{e_{z}}$. In this case it is
possible to obtain an explicit form for the evolution of the linear part, 
\begin{equation}
\left( 
\begin{array}{l}
\stackrel{\rightarrow }{\xi _{1}}\left( t\right) \\ 
\stackrel{\rightarrow }{\xi _{2}}\left( t\right)
\end{array}
\right) =e^{t\left( \stackrel{\rightarrow }{\xi _{1}}\cdot \nabla _{\xi
_{2}}+\frac{e_{i}}{cm_{i}}\nabla _{\xi _{2}}\times \stackrel{\rightarrow }{B}%
_{0}\cdot \stackrel{\rightarrow }{\xi _{2}}\right) }\left( 
\begin{array}{l}
\stackrel{\rightarrow }{\xi _{1}} \\ 
\stackrel{\rightarrow }{\xi _{2}}
\end{array}
\right)  \label{3.5}
\end{equation}
that is 
\begin{equation}
\begin{array}{lll}
\frac{d}{dt}\stackrel{\rightarrow }{\xi }_{1}\left( t\right) & = & 0 \\ 
\frac{d}{dt}\stackrel{\rightarrow }{\xi }_{2}\left( t\right) & = & \stackrel{%
\rightarrow }{\xi _{1}}\left( t\right) +\frac{e_{i}}{cm_{i}}B_{0}\times 
\stackrel{\rightarrow }{\xi _{2}}\left( t\right)
\end{array}
\label{3.6}
\end{equation}
with solution $\stackrel{\rightarrow }{\xi _{1}}\left( t\right) =\stackrel{%
\rightarrow }{\xi _{1}}$ and 
\begin{equation}
\begin{array}{lll}
\left( \xi _{2}\left( t\right) \right) _{x} & = & -\left( \xi _{2}\right)
_{y}\sin \omega _{i}t+\left( \xi _{2}\right) _{x}\cos \omega _{i}t+\frac{1}{%
\omega _{i}}\left\{ \left( \xi _{1}\right) _{x}\sin \omega _{i}t+\left( \xi
_{1}\right) _{y}\left( \cos \omega _{i}t-1\right) \right\} \\ 
\left( \xi _{2}\left( t\right) \right) _{y} & = & \left( \xi _{2}\right)
_{x}\sin \omega _{i}t+\left( \xi _{2}\right) _{y}\cos \omega _{i}t+\frac{1}{%
\omega _{i}}\left\{ \left( \xi _{1}\right) _{x}\left( 1-\cos \omega
_{i}t\right) +\left( \xi _{1}\right) _{y}\sin \omega _{i}t\right\} \\ 
\left( \xi _{2}\left( t\right) \right) _{z} & = & \left( \xi _{2}\right)
_{z}+t\left( \xi _{1}\right) _{z}
\end{array}
\label{3.7}
\end{equation}
with $\omega _{i}=\frac{e_{i}B_{0}}{cm_{i}}$, the inverse relation being 
\begin{equation}
\begin{array}{ccl}
\left( 
\begin{array}{c}
\left( \xi _{2}\right) _{x} \\ 
\left( \xi _{2}\right) _{y}
\end{array}
\right) & = & \left( 
\begin{array}{ll}
\cos \omega _{i}t & \sin \omega _{i}t \\ 
-\sin \omega _{i}t & \cos \omega _{i}t
\end{array}
\right) \left( 
\begin{array}{c}
\left( \xi _{2}\left( t\right) \right) _{x}-\frac{1}{\omega _{i}}\left(
\left( \xi _{1}\right) _{x}\sin \omega _{i}t+\left( \xi _{1}\right)
_{y}\left( \cos \omega _{i}t-1\right) \right) \\ 
\left( \xi _{2}\left( t\right) \right) _{y}-\frac{1}{\omega _{i}}\left\{
\left( \xi _{1}\right) _{x}\left( 1-\cos \omega _{i}t\right) +\left( \xi
_{1}\right) _{y}\sin \omega _{i}t\right\}
\end{array}
\right) \\ 
\left( \xi _{2}\right) _{z} & = & \left( \xi _{2}\left( t\right) \right)
_{z}-t\left( \xi _{1}\right) _{z}
\end{array}
\label{3.7a}
\end{equation}

In integral form Eq.(\ref{3.4}) becomes

\begin{eqnarray}
F_{i}\left( \xi _{1},\xi _{2},t\right)  &=&F_{i}\left( \xi _{1},\xi
_{2}\left( t\right) ,0\right) -\frac{8\pi e_{i}}{m_{i}}\int_{0}^{t}ds\int
d^{3}\xi _{1}^{^{\prime }}F_{i}\left( \xi _{1}-\xi _{1}^{^{\prime }},\xi
_{2}\left( s\right) ,t-s\right)   \nonumber \\
&&\times \frac{\stackrel{\rightarrow }{\xi _{2}}\left( s\right) \cdot 
\stackrel{\rightarrow }{\xi _{1}^{^{\prime }}}}{\left| \xi _{1}^{\prime
}\right| ^{2}}\sum_{j}\frac{1}{2}e_{j}F_{j}\left( \xi _{1}^{^{\prime
}},0,t-s\right)   \label{3.8}
\end{eqnarray}
A stochastic solution is going to be written for the following function 
\begin{equation}
\chi _{i}\left( \xi _{1},\xi _{2},t\right) =e^{-t\gamma \left( \left| \xi
_{2}\right| \right) }\frac{F_{i}\left( \xi _{1},\xi _{2},t\right) }{h\left(
\xi _{1}\right) }  \label{3.8a}
\end{equation}
where $\gamma \left( \left| \xi _{2}\right| \right) =1$ if $\left| \xi
_{2}\right| \leq 1$ and $\gamma \left( \left| \xi _{2}\right| \right)
=\left| \xi _{2}\right| $ otherwise. $h\left( \xi _{1}\right) $ a positive
function to be specified later on. The integral equation for $\chi
_{i}\left( \xi _{1},\xi _{2},t\right) $ is 
\begin{eqnarray}
\chi _{i}\left( \xi _{1},\xi _{2},t\right)  &=&e^{-t\gamma \left( \left| \xi
_{2}\right| \right) }\chi _{i}\left( \xi _{1},\xi _{2}\left( t\right)
,0\right) -\frac{8\pi e_{i}N\left( \xi _{1},\xi _{2},t\right) }{m_{i}}\frac{%
\left( \left| \xi _{1}^{\prime }\right| ^{-1}h*h\right) \left( \xi
_{1}\right) }{h\left( \xi _{1}\right) }  \nonumber \\
&&\times \int_{0}^{t}ds\frac{\gamma \left( \left| \xi _{2}\left( s\right)
\right| \right) }{N\left( \xi _{1},\xi _{2},t\right) }e^{\left( t-s\right)
\gamma \left( \left| \xi _{2}\left( s\right) \right| \right) -t\gamma \left(
\left| \xi _{2}\right| \right) }\int d^{3}\xi _{1}^{^{\prime }}p\left( \xi
_{1},\xi _{1}^{^{\prime }}\right) \chi _{i}\left( \xi _{1}-\xi
_{1}^{^{\prime }},\xi _{2}\left( s\right) ,t-s\right)   \nonumber \\
&&\times \frac{\stackrel{\rightarrow }{\xi _{2}}\left( s\right) \cdot 
\stackrel{\wedge }{\xi _{1}^{^{\prime }}}}{\gamma \left( \left| \xi
_{2}\left( s\right) \right| \right) }\sum_{j}\frac{1}{2}e_{j}\chi _{j}\left(
\xi _{1}^{^{\prime }},0,t-s\right)   \label{3.9}
\end{eqnarray}
with $\stackrel{\wedge }{\xi _{1}^{^{\prime }}}=\xi _{1}^{^{\prime }}/\left|
\xi _{1}^{^{\prime }}\right| $ , 
\begin{equation}
\left( \left| \xi _{1}^{^{\prime }}\right| ^{-1}h*h\right) \left( \xi
_{1}\right) =\int d^{3}\xi _{1}^{^{\prime }}\left| \xi _{1}^{^{\prime
}}\right| ^{-1}h\left( \xi _{1}-\xi _{1}^{^{\prime }}\right) h\left( \xi
_{1}^{^{\prime }}\right)   \label{3.10}
\end{equation}
and 
\begin{equation}
p\left( \xi _{1},\xi _{1}^{^{\prime }}\right) =\frac{\left| \xi
_{1}^{^{\prime }}\right| ^{-1}h\left( \xi _{1}-\xi _{1}^{^{\prime }}\right)
h\left( \xi _{1}^{^{\prime }}\right) }{\left( \left| \xi _{1}^{^{\prime
}}\right| ^{-1}h*h\right) }  \label{3.11}
\end{equation}

Eq.(\ref{3.9}) has a stochastic interpretation as an exponential plus a
branching process. The survival probability up to time $t$ of the
exponential process is 
\begin{equation}
e^{-t\gamma \left( \left| \xi _{2}\right| \right) }  \label{3.11a}
\end{equation}
and $ds\Pi \left( \xi _{1},\xi _{2},s\right) $ is the decay probability in
time $ds$, with 
\begin{equation}
\Pi \left( \xi _{1},\xi _{2},s\right) =\frac{\gamma \left( \left| \xi
_{2}\left( s\right) \right| \right) e^{\left( t-s\right) \gamma \left(
\left| \xi _{2}\left( s\right) \right| \right) -t\gamma \left( \left| \xi
_{2}\right| \right) }}{N\left( \xi _{1},\xi _{2},t\right) }  \label{3.12}
\end{equation}
$N\left( \xi _{1},\xi _{2},t\right) $ being a normalizing function 
\begin{equation}
N\left( \xi _{1},\xi _{2},t\right) =\frac{1}{1-e^{-t\gamma \left( \left| \xi
_{2}\right| \right) }}\int_{0}^{t}ds\gamma \left( \left| \xi _{2}\left(
s\right) \right| \right) e^{\left( t-s\right) \gamma \left( \left| \xi
_{2}\left( s\right) \right| \right) -t\gamma \left( \left| \xi _{2}\right|
\right) }  \label{3.13}
\end{equation}

In the branching process, $p\left( \xi _{1},\xi _{1}^{^{\prime }}\right)
d^{3}\xi _{1}^{^{\prime }}$ is the probability that, from a $\xi _{1}$ mode,
one obtains a $\left( \xi _{1}-\xi _{1}^{^{\prime }},\xi _{1}^{^{\prime
}}\right) $ branching with $\xi _{1}^{^{\prime }}$ in the volume $\left( \xi
_{1}^{^{\prime }},\xi _{1}^{^{\prime }}+d^{3}\xi _{1}^{^{\prime }}\right) $.

The stochastic interpretation of Eq.(\ref{3.9}) provides a way to compute
the solution. $\chi _{i}\left( \xi _{1},\xi _{2},t\right) $ is computed from
the expectation value of a multiplicative functional associated to the
process. Convergence of the multiplicative functional hinges on the
fulfilling of the following conditions :

(A) $\left| \frac{F_{i}\left( \xi _{1},\xi _{2},0\right) }{h\left( \xi
_{1}\right) }\right| \leq 1$

(B) $\left( \left| \xi _{1}^{^{\prime }}\right| ^{-1}h*h\right) \left( \xi
_{1}\right) \leq h\left( \xi _{1}\right) $

Condition (B) is satisfied, for example, for 
\begin{equation}
h\left( \xi _{1}\right) =\frac{c}{\left( 1+\left| \xi _{1}\right|
^{2}\right) ^{2}}\hspace{1cm}\mathnormal{and}\hspace{1cm}c\leq \frac{1}{3\pi 
}  \label{3.14}
\end{equation}
Indeed computing $\left| \xi _{1}^{^{\prime }}\right| ^{-1}h*h$ one obtains 
\begin{equation}
\begin{array}{lll}
c^{2}\Gamma \left( \xi _{1}\right) =\left( \left| \xi _{1}^{^{\prime
}}\right| ^{-1}h*h\right) \left( \xi _{1}\right) & =2\pi c^{2} & \left\{ 
\frac{2\ln \left( 1+\left| \xi _{1}\right| ^{2}\right) }{\left| \xi
_{1}\right| ^{2}\left( \left| \xi _{1}\right| ^{2}+4\right) ^{2}}+\frac{1}{%
\left| \xi _{1}\right| ^{2}\left( \left| \xi _{1}\right| ^{2}+4\right) }%
\right. \\ 
&  & \left. +\frac{\left| \xi _{1}\right| ^{2}-4}{2\left| \xi _{1}\right|
^{3}\left( \left| \xi _{1}\right| ^{2}+4\right) ^{2}}\left( \frac{\pi }{2}%
-\tan ^{-1}\left( \frac{2-2\left| \xi _{1}\right| ^{2}}{4\left| \xi
_{1}\right| }\right) \right) \right\}
\end{array}
\label{3.15}
\end{equation}
Then $\frac{1}{h\left( \xi _{1}\right) }\left( \left| \xi _{1}^{^{\prime
}}\right| ^{-1}h*h\right) \left( \xi _{1}\right) $ is bounded by a constant
for all $\left| \xi _{1}\right| $, and choosing $c$ sufficiently small,
condition (B) is satisfied.

Once $h\left( \xi _{1}\right) $ consistent with (B) is found, condition (A)
only puts restrictions on the initial conditions. Now one constructs the
following backwards-in-time process, denoted $X\left( \xi _{1},\xi
_{2},t\right) $:

Starting at $\left( \xi _{1},\xi _{2},t\right) $, a particle of species $i$
lives for a $\Pi \left( \xi _{1},\xi _{2},s\right) -$distributed time $s$,
up to time $t-s$, with survival and decay probabilities given by (\ref{3.11a}%
) and (\ref{3.12}). At its death a coin $l_{s}$ (probabilities $\frac{1}{2},%
\frac{1}{2}$) is tossed. If $l_{s}=0$ two new particles of the same species
as the original one are born at time $t-s$ with Fourier modes $\left( \xi
_{1}-\xi _{1}^{^{\prime }},\xi _{2}\left( s\right) \right) $ and $\left( \xi
_{1}^{^{\prime }},0\right) $ with probability density $p\left( \xi _{1},\xi
_{1}^{^{\prime }}\right) $. If $l_{s}=1$ the two new particles are of
different species. Each one of the newborn particles continues its
backward-in-time evolution, following the same decay and branching laws.
When one of the particles of this tree reaches time zero it samples the
initial condition. The multiplicative functional of the process is the
product of the following contributions:

- At each branching point where two particles are born, the coupling
constant is 
\begin{equation}
g_{ij}\left( \xi _{1},\xi _{1}^{^{\prime }},s\right) =-\frac{8\pi
e_{i}e_{j}N\left( \xi _{1},\xi _{2},t\right) }{m_{i}}\frac{\left( \left| \xi
_{1}^{^{\prime }}\right| ^{-1}h*h\right) \left( \xi _{1}\right) }{h\left(
\xi _{1}\right) }\frac{\stackrel{\rightarrow }{\xi _{2}}\left( s\right)
\cdot \stackrel{\wedge }{\xi _{1}^{^{\prime }}}}{\gamma \left( \left| \xi
_{2}\left( s\right) \right| \right) }  \label{3.16}
\end{equation}

- When one particle reaches time zero and samples the initial condition the
coupling is 
\begin{equation}
g_{0i}\left( \xi _{1},\xi _{2}\right) =\frac{F_{i}\left( \xi _{1},\xi
_{2},0\right) }{h\left( \xi _{1}\right) }  \label{3.17}
\end{equation}

The multiplicative functional is the product of all these couplings for each
realization of the process $X\left( \xi _{1},\xi _{2},t\right) $. The
solution $\chi _{i}\left( \xi _{1},\xi _{2},t\right) $ is the expectation
value of the multiplicative functional. 
\begin{equation}
\chi _{i}\left( \xi _{1},\xi _{2},t\right) =\mathbb{E}\left\{ \Pi \left(
g_{0}g_{0}^{^{\prime }}\cdots \right) \left( g_{ii}g_{ii}^{^{\prime }}\cdots
\right) \left( g_{ij}g_{ij}^{^{\prime }}\cdots \right) \right\}  \label{3.19}
\end{equation}
Fig.1 illustrates a realization of the process. Notice that the label $%
0\left( s_{3}-s_{1}\right) $ denotes the mode $\xi _{2}=0$ (anti-)evolved
during the time $s_{3}-s_{1}$ according to Eq.(\ref{3.7}).

\begin{figure}[tbh]
\begin{center}
\psfig{figure=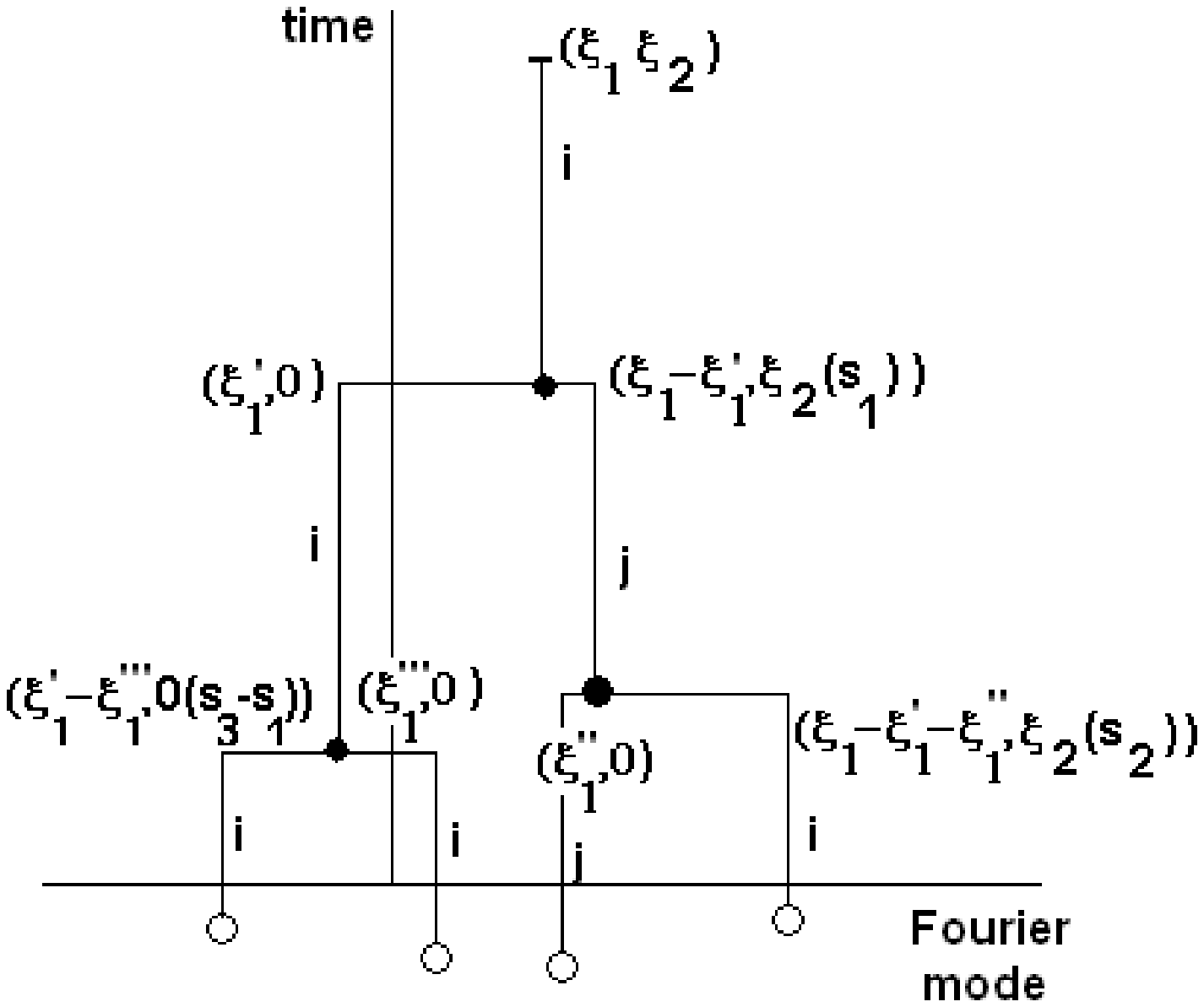,width=11truecm}
\end{center}
\caption{A sample path of the stochastic process $X\left( \xi _{1},\xi
_{2},t\right)$}
\end{figure}

The process itself is the limit of the following iterative process, 
\begin{eqnarray}
&&X_{i}^{\left( k+1\right) }\left( \xi _{1},\xi _{2},t\right)  \nonumber \\
&=&\chi _{i}\left( \xi _{1},\xi _{2}\left( t\right) ,0\right) \mathbf{1}%
_{\left[ s>\tau \right] }+g_{ii}\left( \xi _{1},\xi _{1}^{^{\prime
}},s\right)  \label{3.18} \\
&&\times X_{i}^{\left( k\right) }\left( \xi _{1}-\xi _{1}^{^{\prime }},\xi
_{2}\left( s\right) ,t-s\right) X_{i}^{\left( k\right) }\left( \xi
_{1}^{^{\prime }},0,t-s\right) \mathbf{1}_{\left[ s<\tau \right] }\mathbf{1}%
_{\left[ l_{s}=0\right] }  \nonumber \\
&&+g_{ij}\left( \xi _{1},\xi _{1}^{^{\prime }},s\right) X_{i}^{\left(
k\right) }\left( \xi _{1}-\xi _{1}^{^{\prime }},\xi _{2}\left( s\right)
,t-s\right) X_{j}^{\left( k\right) }\left( \xi _{1}^{^{\prime
}},0,t-s\right) \mathbf{1}_{\left[ s<\tau \right] }\mathbf{1}_{\left[
l_{s}=1\right] }  \nonumber
\end{eqnarray}
with the solution being 
\[
\chi _{i}\left( \xi _{1},\xi _{2},t\right) =\mathbb{E}\left\{
\lim_{k}X_{i}^{\left( k\right) }\left( \xi _{1},\xi _{2},t\right) \right\} 
\]

With the conditions (A) and (B) and choosing the $c$ constant in $h\left(
\xi _{1}\right) $ such that 
\begin{equation}
\left| \frac{8\pi e_{i}e_{j}N\left( \xi _{1},\xi _{2},t\right) }{%
\min_{i}\left\{ m_{i}\right\} }\frac{\left( \left| \xi _{1}^{^{\prime
}}\right| ^{-1}h*h\right) }{h\left( \xi _{1}\right) }\right| \leq 1
\label{3.20}
\end{equation}
the absolute value of all coupling constants is bounded by one. The
branching process, being identical to a Galton-Watson process, terminates
with probability one and the number of inputs to the functional is finite
(with probability one). With the bounds on the coupling constants, the
multiplicative functional is bounded by one in absolute value almost surely.

Once a stochastic solution is obtained for $\chi _{i}\left( \xi _{1},\xi
_{2},t\right) $, one also has, by (\ref{3.8}), a stochastic solution for $%
F_{i}\left( \xi _{1},\xi _{2},t\right) $. Summarizing:

\begin{theorem}
\textit{The stochastic process }$X\left( \xi _{1},\xi _{2},t\right) $\textit{%
, above described, provides through the multiplicative functional (\ref{3.19}%
) a stochastic solution of the Fourier-transformed Poisson-Vlasov equation
in a uniform magnetic field\ for arbitrary finite values of the arguments,
provided the initial conditions at time zero satisfy the boundedness
conditions (A).}
\end{theorem}

\subsection{Fourier-transformed Poisson-Vlasov in a static non-uniform
magnetic field}

The result is now generalized to the case of a static non-uniform magnetic
field. Decompose the Fourier transform of the magnetic field into 
\begin{equation}
\stackrel{\rightarrow }{B}\left( \xi _{1}\right) =\left( 2\pi \right) ^{3/2}%
\stackrel{\rightarrow }{B}_{0}\delta ^{3}\left( \xi _{1}\right) +\stackrel{%
\rightarrow }{b}\left( \xi _{1}\right)   \label{5.1}
\end{equation}
where $\stackrel{\rightarrow }{B}_{0}$ might be the average of the field in
a region of interest and the non-uniform part $\stackrel{\rightarrow }{b}%
\left( \xi _{1}\right) $ is assumed to be small, in a sense to be specified
later. Then the integral equation becomes 
\begin{eqnarray}
F_{i}\left( \xi _{1},\xi _{2},t\right)  &=&F_{i}\left( \xi _{1},\xi
_{2}\left( t\right) ,0\right) -\frac{8\pi e_{i}}{m_{i}}\int_{0}^{t}ds\int
d^{3}\xi _{1}^{^{\prime }}F_{i}\left( \xi _{1}-\xi _{1}^{^{\prime }},\xi
_{2}\left( s\right) ,t-s\right)   \nonumber \\
&&\times \frac{\stackrel{\rightarrow }{\xi _{2}}\left( s\right) \cdot 
\stackrel{\rightarrow }{\xi _{1}^{^{\prime }}}}{\left| \xi _{1}^{\prime
}\right| ^{2}}\sum_{j}\frac{1}{2}e_{j}F_{j}\left( \xi _{1}^{^{\prime
}},0,t-s\right) -\frac{e_{i}}{\left( 2\pi \right) ^{3/2}cm_{i}}%
\int_{0}^{t}ds\int d^{3}\xi _{1}^{^{\prime }}  \nonumber \\
&&\times \stackrel{\rightarrow }{\xi _{2}}\left( s\right) \cdot \left( 
\stackrel{\rightarrow }{b}\left( \xi _{1}^{\prime }\right) \times
\bigtriangledown _{\xi _{2}\left( s\right) }\right) F_{i}\left( \xi _{1}-\xi
_{1}^{^{\prime }},\xi _{2}\left( s\right) ,t-s\right)   \label{5.2}
\end{eqnarray}
where, as before, the dynamics of the arguments $\xi _{2}\left( t\right) $
in $F_{i}$ is controlled by the constant $\stackrel{\rightarrow }{B}_{0}$
(Eqs.(\ref{3.6}) and (\ref{3.7})). A stochastic solution will be obtained
for the function 
\begin{equation}
\chi _{i}\left( \xi _{1},\xi _{2},t\right) =e^{-t\gamma \left( \left| \xi
_{2}\right| \right) }\frac{F_{i}\left( \xi _{1},\xi _{2},t\right) }{h\left(
\xi _{1}\right) }  \label{5.3}
\end{equation}
with integral equation 
\begin{eqnarray}
\chi _{i}\left( \xi _{1},\xi _{2},t\right)  &=&e^{-t\gamma \left( \left| \xi
_{2}\right| \right) }\chi _{i}\left( \xi _{1},\xi _{2}\left( t\right)
,0\right) -\frac{e_{i}N\left( \xi _{1},\xi _{2},t\right) }{m_{i}}\frac{%
\left( \left| \xi _{1}^{\prime }\right| ^{-1}h*h\right) \left( \xi
_{1}\right) }{h\left( \xi _{1}\right) }\int_{0}^{t}ds\frac{\gamma \left(
\left| \xi _{2}\left( s\right) \right| \right) }{N\left( \xi _{1},\xi
_{2},t\right) }  \nonumber \\
&&\times e^{\left( t-s\right) \gamma \left( \left| \xi _{2}\left( s\right)
\right| \right) -t\gamma \left( \left| \xi _{2}\right| \right) }\int
d^{3}\xi _{1}^{^{\prime }}p\left( \xi _{1},\xi _{1}^{^{\prime }}\right)
\left\{ \frac{1}{2}\frac{16\pi \stackrel{\rightarrow }{\xi _{2}}\left(
s\right) \cdot \stackrel{\wedge }{\xi _{1}^{^{\prime }}}}{\gamma \left(
\left| \xi _{2}\left( s\right) \right| \right) }\sum_{j}\frac{1}{2}e_{j}\chi
_{j}\left( \xi _{1}^{^{\prime }},0,t-s\right) \right.   \nonumber \\
&&\left. +\frac{1}{2}\frac{2}{\left( 2\pi \right) ^{3/2}}\frac{\stackrel{%
\rightarrow }{\xi _{2}}\left( s\right) }{\gamma \left( \left| \xi _{2}\left(
s\right) \right| \right) }\cdot \left( \frac{\stackrel{\rightarrow }{b}%
\left( \xi _{1}^{^{\prime }}\right) }{h\left( \xi _{1}^{^{\prime }}\right) }%
\times \bigtriangledown _{\xi _{2}\left( s\right) }\right) \right\} \chi
_{i}\left( \xi _{1}-\xi _{1}^{^{\prime }},\xi _{2}\left( s\right)
,t-s\right)   \label{5.4}
\end{eqnarray}

As before, a backwards-in-time process, rooted at $\left( \xi _{1},\xi
_{2},t\right) $, is considered. The survival and branching probabilities are
also ruled by (\ref{3.11a})\ and (\ref{3.12}). However now, whenever the
propagating particle dies, there are three distinct possibilities. Either
two new particles of the same species (or of opposite species) are born at
time $t-s$ with Fourier modes $\left( \xi _{1}-\xi _{1}^{^{\prime }},\xi
_{2}\left( s\right) \right) $ and $\left( \xi _{1}^{^{\prime }},0\right) $
with probability density $p\left( \xi _{1},\xi _{1}^{^{\prime }}\right) $
given by (\ref{3.11}), or it is just one particle with mode $\left( \xi
_{1}-\xi _{1}^{^{\prime }},\xi _{2}\left( s\right) \right) $ that is born
and the process samples the field $\stackrel{\rightarrow }{b}\left( \xi
_{1}^{^{\prime }}\right) $. That is, the particle samples the non-uniform
field and is scattered by it. This particle also receives an operator label 
\begin{equation}
K\left( \xi _{1}^{^{\prime }},\xi _{2}\left( s\right) \right) =\frac{2}{%
\left( 2\pi \right) ^{3/2}}\frac{\stackrel{\rightarrow }{\xi _{2}}\left(
s\right) }{\gamma \left( \left| \xi _{2}\left( s\right) \right| \right) }%
\cdot \left( \frac{\stackrel{\rightarrow }{b}\left( \xi _{1}^{^{\prime
}}\right) }{h\left( \xi _{1}^{^{\prime }}\right) }\times \bigtriangledown
_{\xi _{2}\left( s\right) }\right)   \label{5.5}
\end{equation}
The operator labels are subsequently inherited by the offspring of this
particle and accumulate until they are finally applied to those offspring
particles that reach time zero. There is no ambiguity in the application of
the operators to the final particles because both the $\xi _{2}\left(
t\right) $ argument of the final particles and the derivatives $%
\bigtriangledown _{\xi _{2}\left( s\right) }$ should be expressed in terms
of the initial $\xi _{1},\xi _{2},\bigtriangledown _{\xi
_{1}},\bigtriangledown _{\xi _{2}}$ using the solutions of Eqs.(\ref{3.4e}-%
\ref{3.4f}).

Denote by $Y\left( \xi _{1},\xi _{2},t\right) $ the process obtained as the
iterative limit of the construction described above. A realization of the
process is illustrated in Fig.2. The boxed $K\left( \xi _{1}^{^{\prime
\prime }},\xi _{2}\left( s_{2}\right) \right) $ denotes the operator label
that is attached to this particle until it (or its progeny) reaches time
zero.

\begin{figure}[tbh]
\begin{center}
\psfig{figure=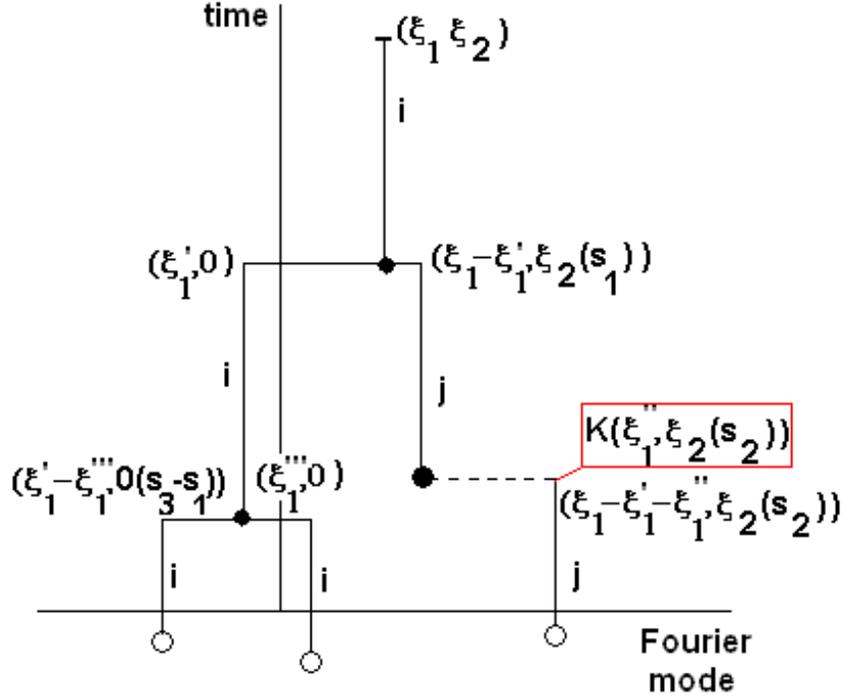,width=11truecm}
\end{center}
\caption{A sample path of the stochastic process $Y\left( \xi _{1},\xi
_{2},t\right)$}
\end{figure}

The solution $\chi _{i}\left( \xi _{1},\xi _{2},t\right) $ of the equation
is then obtained from the average value of a multiplicative functional
associated to the process. For each realization of the process, the
functional is the product of the following factors:

- At each branching point where two particles are born , the coupling
constant is 
\begin{equation}
g_{ij}\left( \xi _{1},\xi _{1}^{^{\prime }},s\right) =-\frac{16\pi
e_{i}e_{j}N\left( \xi _{1},\xi _{2},t\right) }{m_{i}}\frac{\left( \left| \xi
_{1}^{^{\prime }}\right| ^{-1}h*h\right) \left( \xi _{1}\right) }{h\left(
\xi _{1}\right) }\frac{\stackrel{\rightarrow }{\xi _{2}}\left( s\right)
\cdot \stackrel{\wedge }{\xi _{1}^{^{\prime }}}}{\gamma \left( \left| \xi
_{2}\left( s\right) \right| \right) }  \label{5.6}
\end{equation}

- When one particle reaches time zero and samples the initial condition the
coupling is 
\begin{equation}
g_{0i}\left( \xi _{1},\xi _{2}\right) =K\left( \xi _{1}^{^{\prime }},\xi
_{2}\left( s_{1}\right) \right) K\left( \xi _{1}^{^{\prime \prime }},\xi
_{2}\left( s_{2}\right) \right) \cdots K\left( \xi _{1}^{^{(n)}},\xi
_{2}\left( s_{n}\right) \right) \frac{F_{i}\left( \xi _{1},\xi _{2},0\right) 
}{h\left( \xi _{1}\right) }  \label{5.7}
\end{equation}

In addition to condition (B) of the previous subsection, a sufficient
condition for the convergence of the functional is 
\begin{equation}
\left| \frac{16\pi e_{i}e_{j}N\left( \xi _{1},\xi _{2},t\right) }{%
\min_{i}\left\{ m_{i}\right\} }\frac{\left( \left| \xi _{1}^{^{\prime
}}\right| ^{-1}h*h\right) }{h\left( \xi _{1}\right) }\right| \leq 1
\label{5.8}
\end{equation}
and 
\begin{equation}
\left| K\left( \xi _{1}^{^{\prime }},\xi _{2}\left( s_{1}\right) \right)
K\left( \xi _{1}^{^{\prime \prime }},\xi _{2}\left( s_{2}\right) \right)
\cdots K\left( \xi _{1}^{^{(n)}},\xi _{2}\left( s_{n}\right) \right) \frac{%
F_{i}\left( \xi _{1},\xi _{2},0\right) }{h\left( \xi _{1}\right) }\right|
\leq 1  \label{5.9}
\end{equation}
for arbitrary $n$ and arbitrary values of the arguments $\xi _{1}^{^{\prime
}},\xi _{2}\left( s\right) $. This last condition requires boundedness and
smoothness of the initial condition as well as a sufficiently small (as
compared to $h\left( \xi _{1}\right) $) non-uniformity field $\stackrel{%
\rightarrow }{b}\left( \xi _{1}\right) $. Summarizing:

\begin{theorem}
\textit{The stochastic process }$Y\left( \xi _{1},\xi _{2},t\right) $\textit{%
, above described, provides a stochastic solution to the Fourier-transformed
Poisson-Vlasov equation in a static non-uniform magnetic field, provided the
initial conditions at time zero and the non-uniform part of the field
satisfy the condition (\ref{5.9}).}
\end{theorem}

\subsection{The configuration space equation}

Eq.(\ref{3.1}) in integral form is 
\begin{eqnarray}
f_{i}\left( x,v,t\right) &=&e^{-tQ_{\eta }}f_{i}\left( x,v,0\right) -\frac{%
e_{i}}{m_{i}}\int_{0}^{t}dse^{-sQ_{\eta }}\int d^{3}x^{^{\prime
}}\sum_{j}e_{j}\int d^{3}uf_{j}\left( x^{^{\prime }},u,t-s\right)  \nonumber
\\
&&\times \frac{\stackrel{\longrightarrow }{x-x^{^{\prime }}}}{\left|
x-x^{^{\prime }}\right| ^{3}}\cdot \nabla _{v}f_{i}\left( x,v,t-s\right)
\label{6.1}
\end{eqnarray}
or 
\begin{eqnarray}
f_{i}\left( x,v,t\right) &=&f_{i}\left( x\left( t\right) ,v\left( t\right)
,0\right) -\frac{e_{i}}{m_{i}}\int_{0}^{t}ds\int d^{3}x^{^{\prime
}}\sum_{j}e_{j}\int d^{3}uf_{j}\left( x^{^{\prime }},u,t-s\right)  \nonumber
\\
&&\times \frac{\stackrel{\longrightarrow }{x\left( s\right) }-\stackrel{%
\rightarrow }{x^{^{\prime }}}}{\left| x\left( s\right) -x^{^{\prime
}}\right| ^{3}}\cdot \nabla _{v\left( s\right) }f_{i}\left( x\left( s\right)
,v\left( s\right) ,t-s\right)  \label{6.2}
\end{eqnarray}
$x\left( t\right) $, $v\left( t\right) $ being the solutions of (\ref{3.4d})
with $x,v$ as initial conditions.

In the Fourier-transformed equation, division by $\gamma \left( \left| \xi
_{2}\right| \right) $ not only regularizes the velocity gradient as it also,
through multiplication of $F_{i}\left( \xi _{1},\xi _{2},t\right) $ by $%
e^{-t\gamma \left( \left| \xi _{2}\right| \right) }$, introduces a natural
time scale for the exponential process that controls the branching. Here,
because division by $\nabla _{v}$ does not make sense, there is no natural
exponential time scale. One could nevertheless multiply $f_{i}\left(
x,v,t\right) $ by $e^{-\lambda t}$, with $\lambda $ a constant, as in Ref.%
\cite{Vilela3} for the equation without magnetic field. However, because of
the nonlinear nature of the second term in (\ref{6.2}), this introduces
strong limitations on the range of $t$ for which the solution may be
constructed. Here a different procedure will be followed. The price to pay
is that, instead of a simple branching process, one needs a more complex
tree-indexed stochastic process.

Let $G_{i}\left( \stackrel{\rightarrow }{x},\stackrel{\rightarrow }{v}%
,t\right) $ be the function 
\begin{equation}
G_{i}\left( \stackrel{\rightarrow }{x},\stackrel{\rightarrow }{v},t\right) =%
\frac{f_{i}\left( \stackrel{\rightarrow }{x},\stackrel{\rightarrow }{v}%
,t\right) }{\varphi _{i}\left( \stackrel{\rightarrow }{x}\left( t\right) ,%
\stackrel{\rightarrow }{v}\left( t\right) \right) }  \label{6.3}
\end{equation}
the $\varphi \left( \stackrel{\rightarrow }{x},\stackrel{\rightarrow }{v}%
\right) $'s being functions to be specified later and $\left( \stackrel{%
\rightarrow }{x}\left( t\right) ,\stackrel{\rightarrow }{v}\left( t\right)
\right) $ the function arguments (anti-)evolved by (\ref{3.4d}). One obtains
the following integral equation for $G_{i}\left( \stackrel{\rightarrow }{x},%
\stackrel{\rightarrow }{v},t\right) $ 
\begin{eqnarray}
G_{i}\left( \stackrel{\rightarrow }{x},\stackrel{\rightarrow }{v},t\right)
&=&G_{i}\left( \stackrel{\rightarrow }{x}\left( t\right) ,\stackrel{%
\rightarrow }{v}\left( t\right) ,0\right) -2\sum_{j}\frac{1}{2}\frac{%
e_{i}e_{j}}{m_{i}}\int_{0}^{t}dsA_{x,v,t}^{(j)}  \nonumber \\
&&\times \int d^{3}x^{\prime }d^{3}up_{x,v,t}^{(j)}\left( \stackrel{%
\rightarrow }{x^{\prime }},\stackrel{\rightarrow }{u},s\right) G_{j}\left( 
\stackrel{\rightarrow }{x^{\prime }},\stackrel{\rightarrow }{u},t-s\right) 
\widehat{\left( \stackrel{\rightarrow }{x}\left( s\right) -\stackrel{%
\rightarrow }{x^{\prime }}\right) }  \nonumber \\
&&\bullet \frac{1}{\varphi _{i}\left( \stackrel{\rightarrow }{x}\left(
t\right) ,\stackrel{\rightarrow }{v}\left( t\right) \right) }\nabla
_{v\left( s\right) }\varphi _{i}\left( \stackrel{\rightarrow }{x}\left(
t\right) ,\stackrel{\rightarrow }{v}\left( t\right) \right) G_{i}\left( 
\stackrel{\rightarrow }{x}\left( s\right) ,\stackrel{\rightarrow }{v}\left(
s\right) ,t-s\right)  \nonumber \\
&&  \label{6.4}
\end{eqnarray}
with $\widehat{y}=\frac{\stackrel{\rightarrow }{y}}{\left| y\right| }$ and 
\begin{equation}
p_{x,v,t}^{(j)}\left( \stackrel{\rightarrow }{x^{\prime }},\stackrel{%
\rightarrow }{u},s\right) =\frac{1}{A_{x,v,t}^{(j)}}\frac{\varphi _{j}\left( 
\stackrel{\rightarrow }{x^{\prime }}\left( t-s\right) ,\stackrel{\rightarrow 
}{u}\left( t-s\right) \right) }{\left| \stackrel{\rightarrow }{x}\left(
s\right) -\stackrel{\rightarrow }{x^{\prime }}\right| ^{2}}  \label{6.5}
\end{equation}
a probability in the space $[0,t]\times \mathbb{R}^{3}\times \mathbb{R}^{3}$%
, $A_{x,v,t}$ being the normalization constant 
\begin{equation}
A_{x,v,t}^{(j)}=\int_{0}^{t}ds\int \int d^{3}x^{\prime }d^{3}u\frac{\varphi
_{j}\left( \stackrel{\rightarrow }{x^{\prime }}\left( t-s\right) ,\stackrel{%
\rightarrow }{u}\left( t-s\right) \right) }{\left| \stackrel{\rightarrow }{x}%
\left( s\right) -\stackrel{\rightarrow }{x^{\prime }}\right| ^{2}}
\label{6.6}
\end{equation}
One of the simplest choices for the functions $\varphi _{i}\left( \stackrel{%
\rightarrow }{x},\stackrel{\rightarrow }{v}\right) $ would be to make it
proportional to the initial condition 
\begin{equation}
\varphi _{i}\left( \stackrel{\rightarrow }{x},\stackrel{\rightarrow }{v}%
\right) =kf_{i}\left( \stackrel{\rightarrow }{x},\stackrel{\rightarrow }{v}%
,0\right)  \label{6.7}
\end{equation}
Then, the probabilistic interpretation would require finiteness of 
\begin{equation}
A_{x,v,t,s}^{(j)}=\int_{0}^{t}ds\int d^{3}x^{\prime }d^{3}u\frac{%
kf_{j}\left( \stackrel{\rightarrow }{x^{\prime }}\left( t-s\right) ,%
\stackrel{\rightarrow }{u}\left( t-s\right) ,0\right) }{\left| \stackrel{%
\rightarrow }{x}\left( s\right) -\stackrel{\rightarrow }{x^{\prime }}\right|
^{2}}  \label{6.8}
\end{equation}
a quantity that has the nature of a retarded field intensity generated by
the initial condition. However, the general result will be stated without
committing to a particular choice of $\varphi _{i}\left( \stackrel{%
\rightarrow }{x},\stackrel{\rightarrow }{v}\right) $.

From Eq.(\ref{6.4}) one sees that because the term $G_{i}\left( \stackrel{%
\rightarrow }{x}\left( t\right) ,\stackrel{\rightarrow }{v}\left( t\right)
,0\right) $ is not multiplied by a probability factor one cannot simply
interpret the construction of $G_{i}\left( \stackrel{\rightarrow }{x},%
\stackrel{\rightarrow }{v},t\right) $ as importance sampling of the Picard
series. Nevertheless, a probabilistic interpretation may be given through
the following tree-indexed stochastic process $Z\left( \stackrel{\rightarrow 
}{x},\stackrel{\rightarrow }{v},t\right) $:

Rooted at $\left( \stackrel{\rightarrow }{x},\stackrel{\rightarrow }{u}%
,t\right) $, a particle of species $i$ propagates backwards-in-time until a
time $t-s$ when, controlled by the probability $p_{x,v,t}^{(j)}\left( 
\stackrel{\rightarrow }{x^{\prime }},\stackrel{\rightarrow }{u},s\right) $
it gives birth to two new particles. One of them is of the same species $i$
and the other of the same or the opposite species with probability $\frac{1}{%
2}$. The first particle has coordinates $\left( \stackrel{\rightarrow }{x}%
\left( s\right) ,\stackrel{\rightarrow }{v}\left( s\right) \right) $ and the
other coordinates $\left( \stackrel{\rightarrow }{x^{\prime }},\stackrel{%
\rightarrow }{u}\right) $ determined by the probability $p_{x,v,t}^{(j)}%
\left( \stackrel{\rightarrow }{x^{\prime }},\stackrel{\rightarrow }{u}%
,s\right) $. The first particle also receives an operator label 
\begin{equation}
K\left( s\right) =\widehat{\left( \stackrel{\rightarrow }{x}\left( s\right) -%
\stackrel{\rightarrow }{x^{\prime }}\right) }\bullet \frac{1}{\varphi
_{i}\left( \stackrel{\rightarrow }{x}\left( t\right) ,\stackrel{\rightarrow 
}{v}\left( t\right) \right) }\nabla _{v\left( s\right) }\varphi _{i}\left( 
\stackrel{\rightarrow }{x}\left( t\right) ,\stackrel{\rightarrow }{v}\left(
t\right) \right)  \label{6.9}
\end{equation}
to be subsequently applied to all of its offspring. The original particle,
the one that gave birth to the two new ones, does not die and proceeds its
free propagation until time zero. Then, each one of the newly created
particles has an evolution analogous to the progenitor and during the its
evolution the operator labels that they inherit at the birth of each new
pair are accumulated, until they are finally applied to the initial
condition when each one of the particles reaches time zero. A realization of
the process is illustrated in Fig.3. The flags, denoted $s_{1},s_{2},\cdots $%
, stand for the operator labels $K\left( s_{1}\right) ,K\left( s_{2}\right)
,\cdots $.

\begin{figure}[tbh]
\begin{center}
\psfig{figure=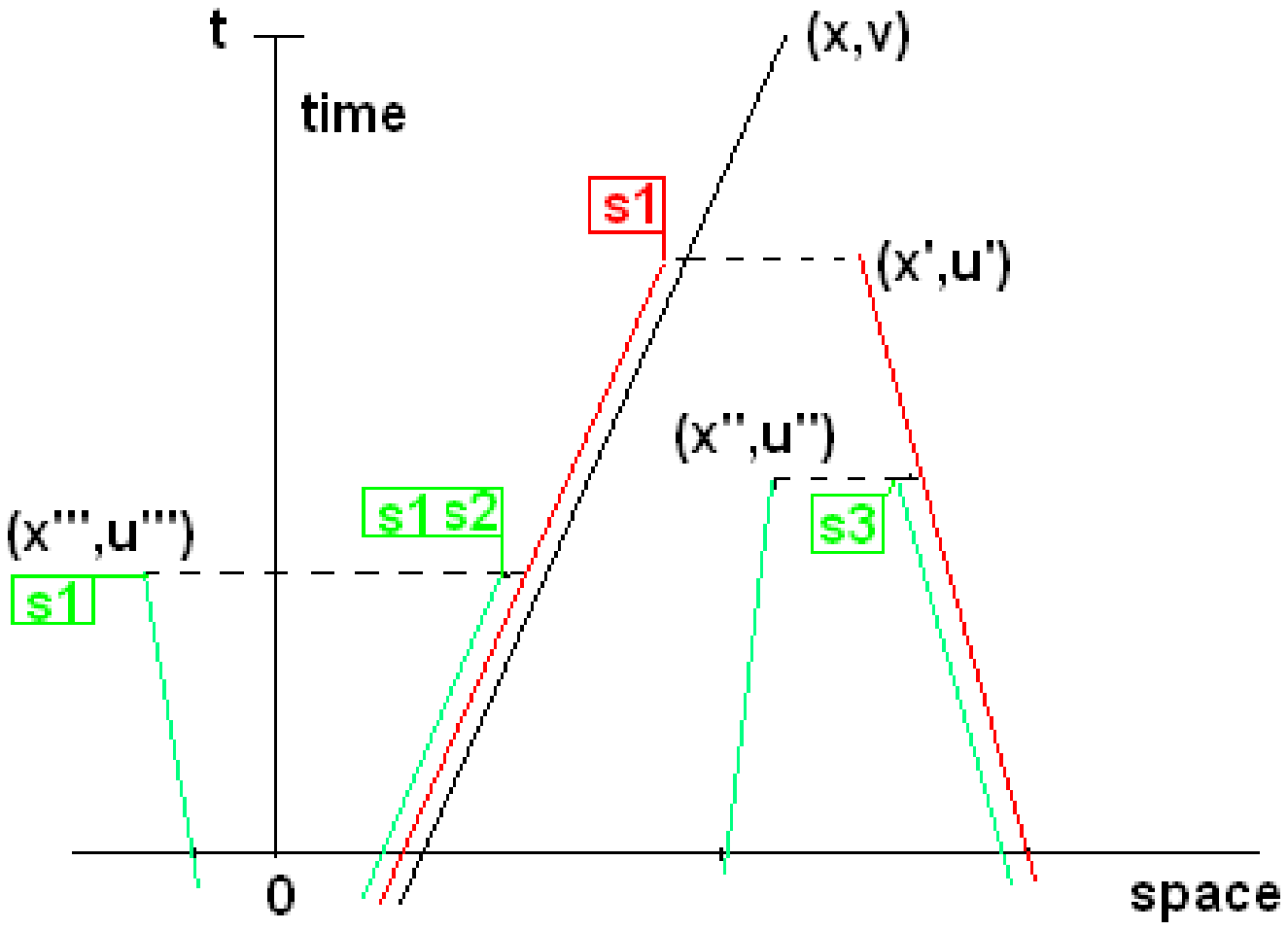,width=11truecm}
\end{center}
\caption{A sample path of the stochastic process $Z\left( \stackrel{%
\rightarrow }{x},\stackrel{\rightarrow }{v},t\right) $}
\end{figure}

The main differences from the Fourier-transformed case are:

- The progenitor particles never die,

- The solution of the equation is obtained from the average over
realizations of the following quantities 
\begin{eqnarray}
\stackrel{\backsim }{G}_{i}\left( \stackrel{\rightarrow }{x},\stackrel{%
\rightarrow }{v},t\right)  &=&G_{i}\left( \stackrel{\rightarrow }{x}\left(
t\right) ,\stackrel{\rightarrow }{v}\left( t\right) ,0\right)   \label{6.10}
\\
&&-\frac{2e_{i}e_{j}}{m_{i}}A_{x,v,t}^{(j)}\stackrel{\backsim }{G}_{j}\left( 
\stackrel{\rightarrow }{x^{\prime }},\stackrel{\rightarrow }{u},t-s\right)
K\left( s\right) \stackrel{\backsim }{G}_{i}\left( \stackrel{\rightarrow }{x}%
\left( s\right) ,\stackrel{\rightarrow }{v}\left( s\right) ,t-s\right)  
\nonumber
\end{eqnarray}
with $\stackrel{\backsim }{G}_{j}\left( \stackrel{\rightarrow }{x^{\prime }},%
\stackrel{\rightarrow }{u},t-s\right) $ and $\stackrel{\backsim }{G}%
_{i}\left( \stackrel{\rightarrow }{x}\left( s\right) ,\stackrel{\rightarrow 
}{v}\left( s\right) ,t-s\right) $ computed in the same way until $t-s$
reaches time zero. For each realization the process runs from time $t$ to
zero. However, the calculation of the quantities $\stackrel{\backsim }{G}_{i}
$ for each realization runs the opposite way, from time zero to time $t$.
Qualitatively, what the process does is to replace the calculation of the
integrals in (\ref{6.4}) by the generation of a family of probability
measures and each value of (\ref{6.10}) is a sampling of the corresponding
Picard iteration.

Assume that, with probability one, the iteration (\ref{6.10}) converges for
all realizations of the process. Then the solution of (\ref{6.4}) is
obtained from 
\begin{equation}
G_{i}\left( \stackrel{\rightarrow }{x},\stackrel{\rightarrow }{v},t\right) =%
\mathbb{E}\left\{ \stackrel{\backsim }{G}_{i}\left( \stackrel{\rightarrow }{x%
},\stackrel{\rightarrow }{v},t\right) \right\}  \label{6.11}
\end{equation}

Hence, existence of the stochastic solution depends on the boundedness and
convergence of the iteration in (\ref{6.10}). Let 
\begin{equation}
\left| G_{i}\left( \stackrel{\rightarrow }{x},\stackrel{\rightarrow }{v}%
,0\right) \right| \leq M  \label{6.12}
\end{equation}
and 
\begin{equation}
\left| K\left( s_{1}\right) K\left( s_{2}\right) \cdots K\left( s_{n}\right)
G_{i}\left( \stackrel{\rightarrow }{x},\stackrel{\rightarrow }{v},0\right)
\right| \leq M  \label{6.13}
\end{equation}
for all $\stackrel{\rightarrow }{x},\stackrel{\rightarrow }{v},n$. Then for
any arbitrary number of steps in the calculation of (\ref{6.10}) one would
obtain a finite value if 
\begin{equation}
8\max \left| \frac{A_{x,v,t}^{(j)}}{m_{i}}\right| M<1  \label{6.14}
\end{equation}

In conclusion:

\begin{theorem}
\textit{If the smoothness and boundedness conditions (\ref{6.12})-(\ref{6.14}%
) are fulfilled, the tree-indexed stochastic process }$Z\left( \stackrel{%
\rightarrow }{x},\stackrel{\rightarrow }{v},t\right) $\textit{\ yields a
stochastic solution of the configuration space Poisson-Vlasov equation in an
external magnetic field.}
\end{theorem}

\section{Remarks and conclusions}

1) The stochastic solution results established for the Fourier-transformed
and the configuration space Poisson-Vlasov equations in an external magnetic
field may, as discussed in the introduction, provide adequate algorithms for
the parallel computation of localized solutions. That implementation of such
algorithms is feasible has been shown in Ref.\cite{Vilela2} for the
stochastic solutions associated to branching processes and multiplicative
functionals. For the Fourier-transformed solutions developed here the
algorithms would be quite similar, the main difference being the slightly
more complex exponential process. However, this extra complexity pays off in
allowing for solutions without upper time bounds.

For the tree-indexed processes, that construct the configuration space
solutions, the implementation could lead to larger computer time
requirements because, for each realization, one has to compute the iteration
in Eq.(\ref{6.10}) and then to average over many realizations.

2) In plasma phenomena in strong magnetic fields there is a hierarchy of
well separated time scales, the Larmor time scale, the bounce time scale and
the drift time scale. Separation of the Larmor time scale led to a beautiful
body of theory that goes by the name of gyrokinetics\cite{Brizard}. A
practical motivation for the gyrokinetics reduction comes from the
possibility to reduce the dimension of the numerical codes from 6 to 5 or 4
dimensions. With the present improvement of multiprocessor computer power
this motivation has somehow become weaker, especially because of the
additional complexity of the gyrokinetics equations if one wants to go
beyond the leading order. That, to obtain any reasonable accuracy, higher
gyrokinetic orders should be included in the numerical calculation, is
indeed to be expected in view of the fact that the exact invariant
associated to the gyrokinetics reduction may, at best, be obtained by a
Borel-summable infinite series\cite{Ghendrih}

Nevertheless, if a reduction of the Larmor time scale is desired, the
stochastic solution approach developed in this paper might also provide an
appropriate framework for this reduction. Notice in particular that, in the
configuration space stochastic solutions, the magnetic field evolution acts
only on the function arguments, that is, on the labels of the stochastic
process not on the process itself. Then, averaging techniques or scalar
function mappings would provide an alternative formulation of gyrokinetics.

3) In the stochastic solutions for the configuration space equation and for
the non-uniform magnetic field case, operator labels associated to the
particles generated in the tree are carried over and applied to the initial
conditions when the process arrives to time zero. This entails some
additional complexity in the calculation of the functionals and in the
smoothness requirements to be imposed on the initial conditions. The need
for these operator labels arises from the singular nature of the propagation
kernels derivatives. A simple (one-dimensional) example illustrates this
point. Let us assume that a probabilistic interpretation is to be given to
an integral containing the factor $\partial _{v}f_{i}\left( x,v,t\right) $.
Then we may replace it by 
\[
-\int \delta ^{\prime }\left( v-v^{\prime }\right) f_{i}\left( x,v^{\prime
},t\right) dv^{\prime } 
\]
but it is not possible to absorb $\delta ^{\prime }\left( v-v^{\prime
}\right) $ into a probability kernel unless some limiting approximation is
used 
\[
\int 2\mathnormal{sign}\left( v-v^{\prime }\right) \left| v-v^{\prime
}\right| \lim_{\varepsilon \rightarrow 0}\sqrt{\frac{1}{\pi \varepsilon ^{3}}%
}e^{-\frac{\left( v-v^{\prime }\right) ^{2}}{\varepsilon }}f_{i}\left(
x,v^{\prime },t\right) 
\]
with sign$\left( v-v^{\prime }\right) $ in the coupling constant and the
rest in the probability kernel. However, the computation of the
approximation entails numerical instabilities and to keep the derivative as
an operator label seems to be a more robust procedure.

A completely different situation occurs if the derivative of the propagation
kernel is smooth. This is the case in the Navier-Stokes equation\cite
{Ossiander}, where by an integration by parts the derivative of the heat
kernel is controlled by a majorizing kernel and absorbed in the probability
measure.

\end{document}